\documentclass[%
reprint,superscriptaddress,
 amsmath,amssymb,
 aps,
]{revtex4-2}

\pdfoutput=1

\usepackage{graphicx}
\usepackage{dcolumn}
\usepackage{bm}
\usepackage{comment}
\usepackage{hyperref}
\usepackage{amsmath}
\usepackage{bbm}
\usepackage{caption}

\newcommand{\hodge}{{\star}}
\newcommand{\dd}{{\rm d}}

\begin{document}

\title{High-Quality Axions from Higher-Form Symmetries in Extra Dimensions}

\author{Nathaniel Craig}
 \email{ncraig@ucsb.edu}
 \affiliation{Department of Physics, University of California, Santa Barbara, CA 93106, USA}
 \affiliation{Kavli Institute for Theoretical Physics, Santa Barbara, CA 93106, USA}
\author{Marius Kongsore}
 \email{mkongsore@nyu.edu}
 \affiliation{Center for Cosmology and Particle Physics, Department of Physics, New York University, New York, NY 10003, USA}

\date{\today}

\begin{abstract}
The extra-dimensional axion solves the strong CP problem while largely circumventing the quality problem that plagues its four-dimensional counterparts. Such high quality can be clearly understood in terms of the generalized global symmetries of the higher-dimensional theory. We emphasize that an electric one-form symmetry is entirely responsible for protecting the potential of axions arising from 5D gauge theories and use this to systematically characterize the extra-dimensional axion quality problem. We identify three ways of breaking this one-form symmetry to generate an axion potential: adding electrically charged matter, gauging a magnetic higher-form symmetry, and turning on an ABJ anomaly. In the latter case, we identify new ways of generating an axion potential via extra-dimensional magnetic monopoles.  We also discuss how the axion is modified if the protective electric one-form symmetry is itself gauged. Finally, we relate these effects to gravitational expectations for the quality problem via generalized weak gravity conjectures. The clarity that generalized symmetries bring to the extra-dimensional axion quality problem highlights their relevance to particle phenomenology.
\end{abstract}

\maketitle

\section{Introduction}

The strong CP problem is one of the great outstanding naturalness puzzles of the Standard Model. Its resistance to anthropic explanation and abundance of future experimental tests make it perhaps the most compelling puzzle at present. Yet solutions to the strong CP problem face their own challenge: they typically require a high-fidelity global symmetry (whether continuous or discrete) at variance with expectations from quantum gravity. This `quality problem'~\cite{Kamionkowski:1992mf, Barr:1992qq, Holman:1992us, Ghigna:1992iv} is particularly acute for the Peccei-Quinn (PQ) axion~\cite{Peccei:1977hh, Peccei:1977ur, Weinberg:1977ma, Wilczek:1977pj}, whose $U(1)_{\rm PQ}$ global symmetry must be protected from Planck-suppressed irrelevant operators to a surprisingly high degree.

The quality problem favors axions arising from the compactification of extra spacetime dimensions, for which the problem is exponentially less severe. Such axions were first encountered in the dimensional reduction of $p$-form fields in string theory~\cite{Witten:1984dg,Choi:1985je,Barr:1985hk,Choi:1985bz,Dine:1986bg,Dine:1986zy,Dine:1986zy,Kallosh:1995hi,Banks:1996ss,Banks:1996ea}, although the essential features can be understood purely in field theory from the reduction of a 5D Abelian gauge field~\cite{Arkani-Hamed:2003xts, Choi_2004}. In these simple field-theoretic models, the axion is related to the zero mode of the gauge field along the extra dimension. Traditionally, the robustness of extra-dimensional axions against the quality problem has been attributed to higher-dimensional gauge invariance, which forbids a local potential for the extra component of the gauge field. While non-local terms involving Wilson loops can be generated from charged fields in the bulk, these are exponentially suppressed by the ratio of the charged mass scale to the compactification scale. In this respect, extra-dimensional axions ``take the log'' of the quality problem~\cite{Reece:2023czb}. However, this explanation is incomplete, as it leverages a redundancy of description (rather than a genuine symmetry), does not account for additional gauge-invariant contributions to the axion potential (such as a Stueckelberg mass for the 5D gauge field), and obscures the connection between the axion's quality and the anticipated violation of global symmetries in a theory of quantum gravity. Select aspects of extra-dimensional axion quality control have recently been studied in refs.~\cite{Burgess:2023ifd,Reece:2024wrn}. However, given the popularity of extra-dimensional axions as a high-quality solution to the strong CP problem, a complete description in terms of global symmetries would be welcome.

Such a description is naturally furnished in the language of generalized global symmetries~\cite{Gaiotto:2014kfa}. In particular, the fate of the extra-dimensional axion's low-energy shift symmetry may be fully understood from the higher-form symmetries of the extra-dimensional gauge theory (as well as their higher-group and non-invertible cousins) \footnote{A satisfying overview of the relevant generalized global symmetries is beyond the scope of this work. For excellent recent reviews germane to the current discussion, see e.g.~\cite{Reece:2023czb,schafernameki2023ictplecturesnoninvertiblegeneralized, Brennan:2023mmt, Bhardwaj:2023kri,Shao:2023gho, Iqbal:2024pee}.}. The relevance of generalized symmetries to extra-dimensional axions is almost as old as the symmetries themselves~\cite{Heidenreich:2015nta, Heidenreich:2015wga, Heidenreich:2020pkc, Cordova:2022rer, Cordova:2022ieu, Reece:2023czb, Reece:2024wrn}. Of particular pertinence to this paper are refs.~\cite{Reece:2023czb,Reece:2024wrn}, which identified a connection between the extra-dimensional axion shift symmetry and a higher-form electric symmetry, and which pointed out that the extra-dimensional axion is a Nambu-Goldstone boson in the same sense that a photon is a Nambu-Goldstone boson of a higher-form symmetry. However, to the best of our knowledge, higher-form symmetries have yet to be used to comprehensively describe the extra-dimensional axion quality problem. In this paper, we enumerate the generalized global symmetries of the simplest extra-dimensional axion arising from the compactification of a 5D Abelian gauge theory, systematically classify the various effects that break these symmetries, and parameterize their impact on extra-dimensional solutions to the strong CP problem. The relative simplicity of this classification underlines the relevance of generalized global symmetries to physics beyond the Standard Model. (See refs.~\cite{Cordova:2022fhg,Cordova:2023her,Aloni:2024jpb,Cordova:2024ypu,Koren:2024xof} for previous applications of generalized symmetries to particle phenomenology.)

We begin in Section~\ref{sec:overview} with a review of extra-dimensional axions from 5D gauge theories, the associated electric 1-form and magnetic 2-form symmetries, and their fates under reduction to 4D. In Section~\ref{sec:electricbreaking} we discuss the three principal ways of generating a potential for the extra-dimensional axion by breaking the electric 1-form symmetry that protects it, and we discuss how the axion is modified if the symmetry is gauged. Although the magnetic 2-form symmetry does not directly protect the axion potential, monopoles can induce effects that break the electric 1-form symmetry, as we describe in Section~\ref{sec:magneticbreaking}. In Section~\ref{sec:gravitationalconsiderations} we relate the breaking of the electric and magnetic higher-form symmetries to the anticipated size of effects arising in a theory of quantum gravity via various weak gravity conjectures and the Completeness Hypothesis. Our conclusions and future directions are summarized in Section~\ref{sec:conclusion}.

\section{Extra-Dimensional Axions and Their Symmetries} \label{sec:overview}
Extra-dimensional axions were first studied in the context of string theory~\cite{Witten:1984dg,Choi:1985je,Barr:1985hk,Choi:1985bz,Dine:1986bg,Dine:1986zy,Dine:1986zy,Kallosh:1995hi,Banks:1996ss,Banks:1996ea}. A general feature of these models is that axions arise as Kaluza-Klein (KK) zero modes of higher-form gauge fields. The behavior of these string theory axions can also be captured by a purely field theoretic model of a one-form gauge field in 5D. In this section, we present a minimal model of this type. We then discuss the higher-form symmetries associated with the extra-dimensional gauge field and how these symmetries connect with axion symmetries in the dimensionally reduced 4D theory.

\subsection{The Axion as a Kaluza-Klein Zero Mode}\label{sec:axionaskkmode}
We adopt a minimal extra-dimensional axion model consisting of a one-form $U(1)$ gauge field $C$ and a one-form $SU(3)$ gauge field $A_G$ existing on an $\mathbb{R}^{1,3}\times S^1$ spacetime~\cite{Arkani-Hamed:2003xts}. One can furthermore add a boundary to the extra-dimensional topology to avoid excess low energy degrees of freedom and for consistent coupling to the Standard Model (see e.g. refs.~\cite{Choi_2004,Grzadkowski:2007xm,Cheng:2001ys} where an orbifold is utilized). We emphasize that these more involved models share the same higher-form symmetry features pointed out in this work~\footnote{For example, if the extra dimension is an $S^1/\mathbb{Z}_2$ orbifold, the survival of the axion zero mode forces us to pick even orbifold parity for $C_5$ so that $C_5(x,-y)=+C_5(x,y)$. This choice poses no obstacle to continuously dragging the symmetry defect operator associated with the electric one-form symmetry safely across the orbifold fixed points, keeping the electric one-form symmetry intact. This in turn rules out the possibility that localized zero mode mass terms can be placed on branes (see also ref.~\cite{vonGersdorff:2002rg} where restrictions on such mass terms are discussed from a rather different perspective).}, but we keep to the simplest possible model as it is most instructive.

We denote the field strength of our $U(1)$ and $SU(3)$ gauge fields by $\dd C$ and $G$, respectively. We adopt the following 5D action
\begin{multline}
    S^\mathrm{5D}=\int\bigg( -\frac{1}{2 g_5^2} \dd C \wedge \star \dd C - \frac{1}{2e_5^2} G \wedge \hodge G \\ + \frac{N}{8\pi^2} C \wedge \mathrm{Tr}\left[G \wedge G \right]\bigg) \, ,
\label{eq:axiongluonaction}
\end{multline}
where $g_5$ and $e_5$ are the 5D gauge couplings of $C$ and $A_G$, respectively, with mass dimension $[g_5]=[e_5]=-1/2$, and where $N\in\mathbb{Z}$. We define the axion as the zero mode of $C$
\begin{equation}
    \theta \equiv \int_{S^1} C \, ,
\label{eq:axiondefinition}
\end{equation}
so that the axion decay constant becomes $f^2\equiv 1/g_5^2 (2\pi R)$. The $2 \pi$ shift symmetry of the axion descends from the invariance of the theory under large gauge transformations $C\rightarrow C+\dd \Lambda_\mathrm{large}$ where $\int\dd\Lambda_\mathrm{large} = 2\pi\mathbb{Z}$ integrated around the circle~\footnote{Note that the term $\dd\Lambda_\mathrm{{large}}$ is somewhat an abuse of notation since the transformations need not be exact. We use this notation to make a distinction between large gauge transformations and the larger class of transformations that define the global one-form symmetry to be discussed in the following section.}. After dimensional reduction, the action~\eqref{eq:axiongluonaction} becomes the standard 4D axion-gluon action
\begin{multline}
    S^\mathrm{4D}= \int\bigg( -\frac{1}{2}f^2 \dd \theta \wedge \hodge \dd \theta - \frac{1}{2 e_4^2}\mathrm{Tr}\,[\tilde{G}\wedge \hodge \tilde{G}]\\
    + \frac{N}{8\pi^2} \theta\, \mathrm{Tr}\,[\tilde{G}\wedge \tilde{G}]+\dots\bigg) \, ,
\label{eq:axionlowenergy}
\end{multline}
where $e_4^2 \equiv e_5^2/(2\pi R)$, where $\tilde{G}$ is the 4D two-form remnant of $G$, and where the dots represent couplings involving higher KK modes, which are irrelevant in the low energy limit of the theory. We have also left out the 4D gauge field component of $C$ and the scalar modes associated with $G$, both of which can be set to zero by a suitable choice of boundary conditions~\cite{Choi_2004,Grzadkowski:2007xm}.

\subsection{Associated Higher-Form Symmetries and their Dimensional Reduction}
\label{sec:higherformsymms}
For now, let us ignore the gluon coupling in~\eqref{eq:axiongluonaction} and consider the higher-form symmetries of the free 5D gauge field $C$. We reintroduce the gluon coupling later in Section~\ref{sec:abjanomaly} in the context of higher-form symmetry breaking. The free gauge field $C$ has a host of associated higher-form symmetries~\cite{Gaiotto:2014kfa,Heidenreich:2020pkc}. Key among these are the electric one-form symmetry $U(1)_e^{(1)}$ and the magnetic two-form symmetry $U(1)_m^{(2)}$. Both symmetries are continuous and have currents $J_e$ and $J_m$, respectively, given by
\begin{equation}
    J_e = \frac{1}{g_5^2}\hodge \dd C \, , \;\;\;\; J_m = \frac{1}{2\pi}\dd C \, ,
\label{eq:5d_currents}
\end{equation}
where we have adopted a convention where the currents of intact symmetries are closed (as opposed to co-closed, as is occasionally seen in the literature). The electric current $J_e$ is conserved due to the 5D equations of motion $\dd J_e\propto\dd \hodge \dd C =0$, while the magnetic current $J_m$ is conserved due to the Bianchi identity $\dd J_m\propto\dd \dd C =0$. We now study the fate of these two symmetries upon dimensional reduction.

In 5D, the transformation associated with the electric one-form symmetry is a shift of the gauge field $C\rightarrow C+\Lambda$ where $\Lambda$ is a flat connection. The $\mathbb{R}^{1,3}$ piece of this shift clearly yields an identical 4D one-form symmetry for the 4D gauge field remnant of $C$, provided this remnant has not been set to zero. Crucially, the $S^1$ piece of the symmetry gives rise to the continuous axion shift symmetry in 4D. This can be seen by simply plugging the shift into the definition of the axion~\eqref{eq:axiondefinition} and carrying out the integral
\begin{equation}
\theta\rightarrow \theta + \int_{S^1} \Lambda = \theta + \mathrm{constant} \, ,
\end{equation}
where the constant takes on values in the entire $[0,2\pi)$ interval. Alternatively, the connection between the one-form symmetry and axion shift symmetry can be seen by noticing that the electric current~\eqref{eq:5d_currents} reduces to the axion shift symmetry current $J_s=f^2\hodge \dd \theta$ upon dimensional reduction.
Hence, the continuous axion zero-form shift symmetry has its higher-dimensional origin in the electric $U(1)_e^{(1)}$ one-form symmetry of $C$ \cite{Reece:2023czb}. This is a crucial observation: the 4D axion potential is protected by the continuous axion shift symmetry. Since this shift symmetry is embedded in a one-form electric symmetry in extra-dimensional models, the quality of the axion can be understood in terms of the breaking and gauging of this electric one-form symmetry. For the axion to acquire a potential, the shift symmetry and hence the electric one-form symmetry of $C$ \textit{must} be broken. But one-form symmetries are ``harder to break'' in the sense that 1) there are generally fewer ways to break them, and 2) explicit breaking has a lesser effect on infrared physics compared to the breaking of ordinary zero-form symmetries. This latter point can be seen in several ways -- e.g. by dimensional analysis, decompactifying an extra dimension, or by appeals to locality~\cite{McGreevy:2022oyu}. It can be more rigorously shown via action deformations in mean string field theory~\cite{Iqbal_2022}. With these preliminary observations out of the way, we emphasize the central lesson of this work:
\begin{center}
    \textit{The continuous axion shift symmetry is embedded in an electric one-form symmetry. This symmetry is difficult to break, protecting the axion potential from dangerous corrections from UV physics.}
\end{center}
This statement should be contrasted with the usual appeal to gauge invariance. The higher-form symmetries perspective has a number of advantages. First, it is a more physical viewpoint since the electric one-form symmetry is a real symmetry of the theory, as opposed to gauge symmetry, which is a redundancy of description. Second, the one-form symmetry breaking viewpoint is on more equal footing with its PQ counterpart. For the PQ axion, the continuous shift symmetry current descends directly from the current associated with the zero-form phase rotation symmetry of a complex scalar field $\phi$, given by $J_p=i\hodge(\phi \dd \phi^* - \phi^* \dd \phi)$~\cite{Peccei:1977hh, Peccei:1977ur, Weinberg:1977ma, Wilczek:1977pj}. Third, the higher-form symmetries viewpoint is much more powerful as an organizing principle. One can write down a host of additional gauge invariant operators in a theory of the form~\eqref{eq:axiongluonaction}, but one finds that only a select few of these operators are capable of generating an axion potential. On the other hand, armed with the knowledge that it is the electric one-form symmetry that protects the axion potential, one only needs to write down the set of operators that break this symmetry. In fact, we find that whenever the symmetry is broken (and cannot be restored via a suitable field or current redefinition), the axion gets a potential, and vice versa. Using this organizing principle, we have naturally been led to all known ways to generate a potential for an extra-dimensional axion of the form~\eqref{eq:axiongluonaction}, which we discuss in Section~\ref{sec:electricbreaking}. We have also been led to new avenues for potential generation, in particular via the interplay between the electric one-form symmetry, magnetic charges, and the magnetic two-form symmetry of $C$.

The 5D magnetic $U(1)_m^{(2)}$ two-form symmetry dimensionally reduces to a magnetic one-form symmetry for the 4D remnant gauge field, provided it has not been set to zero by boundary conditions, and a two-form winding symmetry for the axion. This can be seen by dualizing $C$ to a two-form field $\tilde{C}$. The magnetic two-form symmetry shifts $\tilde{C}$ by a flat connection $\tilde{C}\rightarrow \tilde{C}+\tilde{\Lambda}$. This amounts to a flat connection shift for the 4D remnant of $\tilde{C}$ and a two-form shift for the \textit{dual} axion~\cite{Kallosh:1995hi,Quevedo:1995ep,Burgess:2023ifd,Dvali:2022fdv,Sakhelashvili:2021eid,Choi:2023gin,Platschorre:2024xxp}. Alternatively, it can be seen directly by dimensional reduction that the 5D magnetic current~\eqref{eq:5d_currents} becomes the axion winding symmetry current $J_w=\frac{1}{2\pi}\dd\theta$ in 4D. While the magnetic two-form symmetry and the symmetries that descend from it play no direct role in protecting the axion potential, their gauging or breaking nevertheless has an important connection with the electric-one form symmetry as discussed in Sections~\ref{sec:gaugingmag} and~\ref{sec:magneticbreaking}.

\section{Modifying the Electric One-Form Symmetry}\label{sec:electricbreaking}
In this section, we discuss the four principal ways of modifying the electric one-form current in the minimal extra-dimensional axion model~\eqref{eq:axiongluonaction} and generating a potential for the axion. We identify three ways of breaking the symmetry: introducing electrically charged matter, gauging a magnetic two-form symmetry, and turning on an ABJ anomaly. We also discuss what happens when the electric one-form symmetry is gauged. These four current modifications and their effects are summarized in Table~\ref{table:symmetrymodification}. Finally, we discuss kinetic mixing and Euler-Heisenberg-type irrelevant operators and how they fail to affect the electric one-form symmetry.

\begin{table*}
\renewcommand{\arraystretch}{1.8}
\begin{tabular}{ | c || c | c | c | } 
  \hline
    Symmetry Modification & Current Equation & Remnant Symmetry & Potential \\ 
  \hline\hline
  Electric Matter & $\dd J_e = j_\mathrm{matter}$ & $ \mathbb{Z}_q^{(1)}$ & $V_\theta \simeq -\frac{(m_\mathrm{5D}R)^2}{(2\pi R)^4}e^{-2\pi R m_\mathrm{5D}}\cos(q\theta) $\\
  \hline
  Magnetic Gauging & $\dd J_e = \frac{M}{2\pi}\dd K$ & $\mathbb{Z}_M^{(1)} $ & $m_\theta = e_{K,4}\, \frac{M}{2\pi f}$\\
  \hline
  Electric Gauging & $kJ_e' = \frac{1}{e_B^2}\dd\hodge \dd B$ & $U(1)_e^{(1)}$ & $m_{\tilde{A}} = k\, e_{B,4}\, f$\\ 
  \hline
  ABJ Anomaly & $\dd J_e = \frac{N}{8
  \pi^2} \mathrm{Tr}\,[\tilde{G} \wedge \tilde{G}] $&  $\mathbb{Z}_N^{(1)}$ & $V_\theta\simeq -\Lambda_{\tilde{G}}^4\cos (N\theta)$ \\ 
  \hline
    \end{tabular}
\caption{The four principal ways of modifying the electric $U(1)_e^{(1)}$ one-form symmetry of the 5D axion gauge field $C$. These are: (1) explicit breaking by electrically charged matter, discussed in Section~\ref{sec:electriccharge}, (2) explicit breaking by gauging a magnetic two-form symmetry, discussed in Section~\ref{sec:gaugingmag} (3), gauging the electric one-form symmetry, discussed in Section~\ref{sec:gaugingelec}, and (4) turning on an ABJ anomaly, discussed in Section~\ref{sec:abjanomaly}. Definitions of fields and constants are found in each respective section. For each type of one-form symmetry modification, the table shows how the current equation $\dd J_e=0$ is modified, what (invertible) subgroup symmetry survives the modification, and what the resultant axion potential is. Additional potential contributions from monopoles, discussed in Section~\ref{sec:magneticbreaking}, fall under these four categories as well. Kinetic mixing terms and Euler-Heisenberg-type irrelevant operators do not break the one-form symmetry, as discussed in Section~\ref{sec:mixingandirrelevant}.}
\label{table:symmetrymodification}
\end{table*}

\subsection{Electrically Charged Matter}\label{sec:electriccharge}
Adding an electrically charged particle is the most straightforward way of breaking the electric $U(1)_e^{(1)}$ one-form symmetry. If the charged particle has field $\phi$ and carries electric charge $q\in \mathbb{Z}$ under $C$, the continuous $U(1)_e^{(1)}$ symmetry breaks to a discrete $\mathbb{Z}_q^{(1)}$ one-form symmetry due to $\phi$ operator insertions being able to completely screen charge $nq$ Wilson lines, where $n\in \mathbb{Z}$~\cite{Gaiotto:2014kfa}. With the introduction of the electrically charged particle, the $U(1)_{e}^{(1)}$ current equation takes on the simple form
\begin{equation}
    \dd J_e = j_\mathrm{matter} \, ,
\end{equation}
where $j_\mathrm{matter}$ is the electric flux source on the right hand side of Gauss' law. This breaking gives rise to an effective potential for the axion via worldline instantons. This phenomenon has been well-studied in refs.~\cite{Hosotani:1983xw,Cheng:2002iz,Arkani-Hamed:2007ryu,Reece:2024wrn}. 

Following ref.~\cite{Reece:2023czb}, a heuristic way of understanding how the potential arises is to realize that each integer winding of the charged particle worldline $\gamma$ around the compact dimension corresponds to a topologically distinct configuration of the underlying theory. The Euclidean action of each of these wrapped worldlines is then
\begin{align}
    S_E &= m_\mathrm{5D}\int_\gamma \dd \tau+i q \int_\gamma C \, ,\\
    &\simeq 2\pi R \omega m_{5D}+i q \omega \theta \, ,
\end{align}
where $m_{\mathrm{5D}}$ is the particle mass and $\omega$ is the worldline winding number. One can then estimate the effective potential as a sum over all worldline instantons
\begin{align}
    V(\theta)&\propto \sum_{\omega\in\mathbb{N}} e^{-S_E(\omega)}+ e^{-S_E(-\omega)} \, ,\\
    &\propto \sum_{\omega\in\mathbb{N}} e^{-2\pi R \omega m_\mathrm{5D}}\cos{(\omega q \theta)} \, .
\end{align}
This rough estimate shows the direct role that the instanton action plays in exponentially suppressing the effective axion potential. From the electric one-form symmetry point of view, where Wilson lines are the appropriate charge operators, it is the non-contractibility of these wrapped worldlines that is responsible for breaking the one-form symmetry badly enough for there to be an effect on the axion potential in the infrared. Their extended nature means they must come exponentially suppressed, and the fact that they can be completely self-enclosing when wrapped along the extra dimension is exactly the statement that the charge operators associated with a one-form symmetry in 5D become the charge operators of a zero-form symmetry in 4D. Note also the $2\pi/q$ periodicity of the potential, reflecting the surviving discrete $\mathbb{Z}_q^{(1)}$ one-form symmetry, which becomes a discrete $\mathbb{Z}_q^{(0)}$ zero-form shift symmetry for the axion in the infrared.

A more careful calculation of the potential, carried out in ref.~\cite{Arkani-Hamed:2007ryu}, involves computing the Casimir energy of the theory using the fact that the 4D propagator must be a sum over all possible 5D propagator windings along the hidden extra dimension. Using the Casimir energy, one can compute the effective axion potential at one-loop order exactly
\begin{equation}
    V(\theta)=-2\pi R\sum_{n=1}^\infty \frac{2 m_\mathrm{5D}^5}{(2\pi)^{5/2}}\frac{K_{5/2}(2\pi R m_\mathrm{5D}n)}{(2\pi R m_\mathrm{5D}n)^{5/2}}\cos(nq\theta) \, ,
\label{eq:electricmatterpotential}
\end{equation}
where $K_{5/2}$ is a modified Bessel function of the second kind~\footnote{Here, we have assumed that the wrapped particle is a scalar.}. In the limit $m_\mathrm{5D}\gg 1/R$, the potential simplifies to
\begin{equation} \label{eq:chargedpot}
    V(\theta)\simeq -\frac{(m_\mathrm{5D}R)^2}{(2\pi R)^4}e^{-2\pi R m_\mathrm{5D}}\cos(q\theta) \, ,
\end{equation}
matching the leading $\omega=1$ term in the simple estimate. This exponential suppression means that heavy particles in our theory obeying $m_\mathrm{5D}\gg 1/R$ only marginally affect the axion potential. This suppression is the origin of the statement that extra-dimensional axions ``take the log'' of the quality problem~\cite{Reece:2023czb}.

\subsection{Gauging the Magnetic Two-Form Symmetry}\label{sec:gaugingmag}
The electric one-form symmetry may also be broken by gauging the magnetic two-form symmetry associated with $C$. To do this, we introduce a three-form gauge field $K$ that couples to $C$ via the addition
\begin{equation}
    S^\mathrm{5D}\supset \int\left(-\frac{1}{2e_K^2}\dd K \wedge \hodge \dd K + \frac{M}{2\pi} K\wedge \dd C\right) \, ,
\label{eq:bf_action}
\end{equation}
where $e_K$ is the three-form gauge coupling with mass dimension $[e_K]=3/2$ and $M\in\mathbb{Z}$. The integer quantization is necessitated by the action only being allowed to shift by a $2\pi\mathbb{Z}$ phase under simultaneous large gauge transformations $C\rightarrow C + \dd \Lambda_\mathrm{large}$ and $K\rightarrow K + \dd \Lambda_\mathrm{large}'$ where $\int \dd \Lambda_\mathrm{large}=\int \dd \Lambda_\mathrm{large}'=2\pi\mathbb{Z}$ integrated over a compact manifold~\cite{Brennan:2023mmt}. Using the equations of motion, the electric current equation becomes
\begin{equation}
\dd J_e = \frac{M}{2\pi}\dd K \, ,
\end{equation}
showing that the electric $U(1)^{(1)}_e$ one-form symmetry is broken. Naively, one could attempt to define a refined current $J_e'=\frac{1}{g_5^2}\hodge \dd C - \frac{M}{2\pi}K$, which is formally conserved. However, the associated symmetry defect operator given by $U(\Sigma)=\exp\left[i\alpha \int_\Sigma J_e'\right]$ supported on an arbitrary codimension two manifold $\Sigma$ is only invariant under a large gauge transformation $K\rightarrow K +\dd \Lambda_{\mathrm{large}}'$ if $\alpha=2\pi\mathbb{Z}/ M$. This shows that the electric $U(1)^{(1)}_e$ symmetry is broken down to a $\mathbb{Z}_M^{(1)}$ symmetry, under which $C$ transforms as $C\rightarrow C + \Lambda_{\mathbb{Z}_{M}}$ where $\Lambda_{\mathbb{Z}_M}$ is closed and integrates to $2\pi\mathbb{Z}/M$ on $S^1$.

The coupling term in eq.~\eqref{eq:bf_action} is the 5D version of the well-studied topological BF action~\cite{Horowitz:1989ng,Banks_2011}. A special feature of this type of theory is that its equations of motion are integrable. The equation of motion for $K$ is
\begin{equation}
    \dd \hodge \dd K = e_K^2 \frac{M}{2\pi} \dd C \, .
\end{equation}
We integrate this equation and substitute the new expression for $\dd K$ into the equation of motion for $C$, yielding
\begin{equation}
    \dd\hodge \dd C = g_5^2 e_K^2 \frac{M^2}{4\pi^2} \hodge (C-\alpha) \, ,
\label{eq:c_mass}
\end{equation}
where $\alpha$ is an arbitrary closed one-form, reflecting the gauge symmetry of $C$. To make the surviving discrete $\mathbb{Z}_M^{(1)}$ symmetry explicit, $\alpha$ can be taken to transform as $\alpha\rightarrow \alpha + \Lambda_{\mathbb{Z}_{M}}$ under shifts of $C$. We see that $C$ acquires a mass $m_C=g_5e_K M / 2\pi$. This is hardly a surprise: if one were to dualize $K$ in eq.~\eqref{eq:bf_action} to a scalar field, one would have found a Stueckelberg action where $C$ ``eats'' the dual $K$. This means that eq.~\eqref{eq:bf_action} can be understood as the Stueckelberg limit of an Abelian Higgs model for the gauge field $C$~\cite{Banks_2011}.

By direct dimensional reduction, we find that eq.~\eqref{eq:c_mass} becomes a mass equation for the axion, yielding $m_\theta=e_{K,4}M/2\pi f$ where $e_{K,4}^2\equiv e_K^2/2\pi R$. Another way to understand this mass is to directly dimensionally reduce~\eqref{eq:bf_action}, yielding the 4D axion action
\begin{equation}
    S^{\mathrm{4D}}\supset\int \bigg(-\frac{1}{2e_{K,4}^2}\dd \tilde{K} \wedge \hodge \dd \tilde{K} +\frac{M}{2\pi} \tilde{K} \wedge \dd\theta +\dots \bigg) \, ,
    \label{eq:4d_bf_action}
\end{equation}
where $\tilde{K}$ is the 4D three-form remnant of $K$, which has no propagating degrees of freedom. The dots here indicate terms involving other KK modes of $C$ and $K$ which do not couple to the axion. An action of the form~\eqref{eq:4d_bf_action} has been studied extensively in the literature, especially in the context of inflation~\cite{Dvali:2001sm,Dvali:2005an,Kaloper:2008qs,Silverstein:2008sg,McAllister_2010,Kaloper:2011jz,Marchesano:2014mla,Heidenreich:2020pkc,Aloni:2024jpb}. Like for the 5D theory, one can integrate out $\tilde{K}$, yielding an axion equation of motion
\begin{equation}
    \dd \hodge \dd \theta = \frac{e_{K,4}^2}{f^2}\frac{M^2}{4\pi^2}\hodge(\theta-\kappa) \, ,
\end{equation}
where $\kappa$ is an arbitrary closed zero-form (which by definition is just an ordinary integration constant). This once again shows that the axion obtains a mass $m_\theta=e_{K,4}M/2\pi f$. Furthermore, the theory enjoys a surviving discrete $\mathbb{Z}^{(0)}_M$ zero-form symmetry, descending from the unbroken $\mathbb{Z}^{(1)}_M$ symmetry in the UV. To make this manifest, $\kappa$ can be taken to transform as $\kappa \rightarrow \kappa + 2\pi\mathbb{Z}/M$ under the axion shift symmetry.

The reason the shift symmetry is not manifest in the equations of motion without imposing a condition on $\kappa$ is rich and has been studied extensively in the context of string theory and inflation~\cite{Kaloper:2008qs,Silverstein:2008sg,McAllister_2010,Kaloper:2011jz,Marchesano:2014mla,Aloni:2024jpb}. While the action~\eqref{eq:4d_bf_action} enjoys a discrete shift symmetry, the axion is monodromatic. Combined with the three-form $\tilde{K}$, the axion effectively decompactifies in field space, and the axion potential becomes multi-branched, with each branch allowing $\theta$ to vary over $\mathbb{R}$. Rather than bringing axion states back to themselves, the axion shift symmetry permutes all of these potential branches, leaving the overall theory invariant. For this reason, this type of axion mass is sometimes referred to as a \textit{monodromy mass}.

Unlike the wrapped electric matter mass discussed in Section~\ref{sec:electriccharge}, this tree-level mass can cause difficulty for the quality of the axion since the mass need not be aligned with the QCD potential. This could be overcome by tuning the dimensionful coupling $e_{K,4}$ to be very small, by not including three-form fields in the 5D theory, or by simply assuming $M=0$, which is a perfectly natural possibility~\cite{Reece:2024wrn}.

\subsection{Gauging the Electric One-Form Symmetry} \label{sec:gaugingelec}
We now study what happens when we gauge the electric $U(1)_e^{(1)}$ one-form symmetry. Gauging renders symmetries exact. This means that the electric one-form symmetry cannot be broken. Hence, one might expect that the axion does not acquire a mass. This expectation is technically correct, although the axion still gets a mass in a different sense: it is ``eaten'' by a gauge field via a Stueckelberg mechanism.

To see this, we gauge $J_e$ by introducing a dynamical two-form gauge field $B$ into our 5D action. We furthermore demand that $B$ couples to the symmetry current $J_e$. Making the coupling gauge invariant requires adopting a Stueckelberg-type action
\begin{multline}\label{eq:stueckelberg_5d}
    S^{5D}=\int \bigg(-\frac{1}{2e_B^2}\dd B \wedge \hodge \dd B \\
    -\frac{1}{2g_5^2}(\dd C- kB)\wedge \hodge (\dd C- kB)\bigg) \, ,
\end{multline}
where $k\in \mathbb{Z}$ and $e_B$ is the 5D $B$ gauge coupling with mass dimension $[e_B]=1/2$. The $U(1)^{(1)}_e$ symmetry has now been promoted to a local symmetry. The action~\eqref{eq:stueckelberg_5d} is invariant under the combined gauge transformation
\begin{equation}
    C\rightarrow C+k\Lambda \, , \;\;\;\; B\rightarrow B + \dd \Lambda \, ,
\end{equation}
where $\Lambda=\Lambda(x)$ is a one-form that can now be taken to depend explicitly on the spacetime coordinate $x$, and which is no longer required to be closed. Taking $\Lambda$ to be a flat connection recovers the global electric one-form symmetry. To ensure gauge invariance, we need to define the refined current
\begin{equation}
    J_e'= \frac{1}{g_5^2}\hodge \dd C - \frac{k}{g_5^2}\hodge B \, ,
\end{equation}
which is manifestly gauge invariant and reflects the Stueckelberg combination of $\dd C$ and $B$ in~\eqref{eq:stueckelberg_5d}. This current is exact, which can be seen from the equation of motion for $B$
\begin{equation}
   k J_e' = \frac{1}{e_B^2}\dd \hodge \dd B \, ,
\label{eq:electriccurrentexactobscured}
\end{equation}
in line with expectations. Exactness also means that the current is conserved $\dd J_e'=0$ as required for any gauge symmetry current.

Although they look very different, the action~\eqref{eq:stueckelberg_5d} and the magnetically gauged action~\eqref{eq:bf_action} are very similar. If we were to dualize $C$ to its two-form dual field, we would recover a BF-type action nearly identical to~\eqref{eq:bf_action}, but where the \textit{dual} axion would have its magnetic one-form symmetry gauged~\footnote{This type of coupling for the dual axion was recently studied by ref.~\cite{Platschorre:2024xxp}.}. In the dual frame, electric symmetries become magnetic and vice versa -- hence all breaking or gauging of the axion gauge field's electric symmetry can equivalently be understood as the gauging or breaking of a magnetic symmetry of the dual field.

The action~\eqref{eq:stueckelberg_5d} is the two-form equivalent of a standard Stueckelberg action for a one-form gauge field~\cite{Ruegg:2003ps}. The two-form $B$ ``eats'' $C$, acquiring a mass $m_B=k e_B/g_5$ via the $B^2$ term. After dimensional reduction, it takes on the more familiar form
\begin{multline}
    S^{4D}= \int\bigg(-\frac{1}{2e_{B,4}^2}\dd \tilde{A} \wedge \hodge \dd \tilde{A}\\
    -\frac{1}{2}f^2(\dd \theta - k \tilde{A})\wedge \hodge (\dd\theta - k\tilde{A})+\dots\bigg) \, ,
\label{eq:4dstueckel}
\end{multline}
where $\tilde{A}$ is the one-form zero mode of $B$, $e_{B,4}^2\equiv e_B^2 (2\pi R)$, and where the dots signify other KK modes in the theory which do not couple to the axion. In the dimensionally reduced description, it is the shift symmetry of the axion that is gauged. The axion is eaten by a one-form gauge field, which acquires mass $m_{\tilde{A}}=k e_{B,4}f$. So while the axion technically remains massless (no $\theta^2$ term appears in the action) and still enjoys a gauged shift symmetry, it is absorbed into the massive gauge field $\tilde{A}$ and can be removed from the action altogether via gauge fixing~\cite{Ruegg:2003ps}.

This type of mass presents a challenge to the axion solution to the strong CP problem for two reasons. First, $\tilde{A}$ acquires a large mass under natural assumptions for $f$ and $e_{B,4}$. Second, when the axion shift symmetry is gauged, it cannot also be explicitly broken without spoiling gauge invariance. This naively makes it difficult to couple the Stueckelberg axion to gluons. However, this limitation can be overcome by introducing additional ingredients (e.g.~the Green–Schwarz mechanism~\cite{Green:1984sg}) to cancel the anomalous variation of the axion-gluon coupling.

\subsection{Breaking by ABJ Anomaly}\label{sec:abjanomaly}
We finally turn our attention back to the gauge field-gluon coupling introduced in Section~\ref{sec:axionaskkmode}, responsible for generating the canonical QCD axion potential. The non-Abelian piece of this action takes the form
\begin{equation}
    S^\mathrm{5D}\supset \int\bigg(- \frac{1}{2e_5^2} G \wedge \hodge G + \frac{N}{8\pi^2} C \wedge \mathrm{Tr}\left[G \wedge G \right]\bigg) \, ,
\label{eq:su(N)_abj_anomaly_action}
\end{equation}
where $G$ is the field strength associated with an $SU(N)$ gauge field $A_G$ with gauge coupling $e_G$ with mass dimension $[e_G]=-1/2$, and where $N\in \mathbb{Z}$ so that the action is invariant under simultaneous large gauge transformations of $C$ and $A_G$. This action breaks the electric $U(1)_e^{(1)}$ one-form symmetry via the equation of motion for $C$
\begin{equation}
    \dd J_e = \frac{N}{8\pi^2} \mathrm{Tr}\left[G\wedge G\right] \, .
\end{equation}
This current equation is the 5D equivalent of a 4D non-Abelian ABJ anomaly equation~\cite{Adler:1969gk,Bell:1969ts}. Like for the magnetically gauged action discussed in Section~\ref{sec:gaugingmag}, one might attempt to define a refined conserved current $J_e' =  J_e - \frac{N}{8\pi^2}\mathrm{Tr}\left[A_G\wedge G+\frac{2}{3} A_G\wedge A_G\wedge A_G\right]$, which is formally conserved, but which is not gauge invariant and hence not a genuine operator of the theory. However, the associated symmetry defect operator given by $U(\Sigma)=\exp\left[i\alpha \int_\Sigma J_e'\right]$ \textit{is} gauge invariant under both small and large gauge transformations as long as $\alpha = 2\pi\mathbb{Z}/N$. This shows that the ABJ action~\eqref{eq:su(N)_abj_anomaly_action} breaks the electric $U(1)^{(1)}_e$ one-form symmetry of $C$ down to a discrete $\mathbb{Z}_N^{(1)}$ one-form symmetry.

As discussed in Section~\ref{sec:axionaskkmode}, the action~\eqref{eq:su(N)_abj_anomaly_action} takes on the familiar 4D QCD axion form upon dimensional reduction
\begin{equation}\label{eq:4d_abj_action}
    S^\mathrm{4D}\supset \int \bigg( - \frac{1}{2 e_4^2}\mathrm{Tr}\,[\tilde{G}\wedge \hodge \tilde{G}]\\
    + \frac{N}{8\pi^2} \theta\, \mathrm{Tr}\,[\tilde{G}\wedge \tilde{G}]+\dots\bigg) \, ,
\end{equation}
where $\tilde{G}$ is the field strength of the 4D one-form remnant of $A_G$ with gauge coupling $e_4^2\equiv e_5^2/(2\pi R)$. Here, the dots indicate other KK modes that do not couple to the axion. The action~\eqref{eq:4d_abj_action} hosts $SU(N)$ instantons. For the standard $SU(3)$ gluon coupling, these instantons generate the textbook QCD axion potential~\cite{Choi:2020rgn}
\begin{equation}
    V(\theta)\simeq -\Lambda_\mathrm{QCD}^4 \cos(N\theta) \, ,
\end{equation}
where the approximately equal symbol emphasizes the fact that the QCD scale prefactor $\Lambda_\mathrm{QCD}^4$ is only approximate. Note that the appearance of the integer $N$, sometimes called the \textit{domain wall number}, in the axion potential reflects the fact that the surviving discrete one-form $\mathbb{Z}_N^{(1)}$ symmetry of $C$ becomes a discrete zero-form $\mathbb{Z}_N^{(0)}$ shift symmetry of the axion upon dimensional reduction.

One might wonder whether adding terms of the form~\eqref{eq:su(N)_abj_anomaly_action}, but with $G$ replaced with a $U(1)$ field strength could generate a potential for the axion -- after all, such an action would break the electric $U(1)_e^{(1)}$ one-form symmetry. However, there are no $U(1)$ instantons on a topologically trivial 4D spacetime, hence there cannot be instantons in the infrared regime of our theory~\cite{Srednicki:2007qs,Shifman:2012zz}. Nevertheless, if the theory contains magnetic monopoles, in line with what is mandated by the magnetic Weak Gravity Conjecture, closed monopole worldvolumes do indeed generate a potential via~\eqref{eq:su(N)_abj_anomaly_action}~\cite{Arkani-Hamed:2006emk,Arkani-Hamed:2007ryu,delaFuente:2014aca,Fan:2021ntg}. We discuss this contribution in Section~\ref{sec:magneticbreaking} and other generalized Weak Gravity Conjecture constraints in Section~\ref{sec:gravitationalconsiderations}.

\subsection{A Note on Kinetic Mixing and Irrelevant Operators}\label{sec:mixingandirrelevant}
In addition to the symmetry breaking mechanisms discussed thus far, the conservation of the electric one-form symmetry current $J_e$ is also violated by the presence of kinetic mixing terms and Euler-Heisenberg type higher-dimensional operator terms. However, these violations can always be removed via current redefinitions. We now discuss each type of term in turn.

We may add a kinetic mixing term of the form~\cite{Holdom:1985ag}
\begin{equation}
    S^{\mathrm{5D}}\supset\int \bigg( - \frac{1}{2e_A^2}\dd A \wedge \hodge \dd A +\frac{\kappa}{g_5 e_A}\dd C \wedge \hodge \dd A \, \bigg),
\label{eq:kineticmixing}
\end{equation}
where $A$ is a new dynamical $U(1)$ one-form gauge field with 5D gauge coupling $e_A$ with mass dimension $[e_A]=-1/2$, and where $\kappa$ is a dimensionless parameter controlling the amount of mixing. A term of the form~\eqref{eq:kineticmixing} modifies the $U(1)_e^{(1)}$ current via the new equation of motion $\frac{1}{g_5^2}\dd \hodge \dd C = \frac{\kappa}{g_5 e_A}\dd \hodge \dd A$. It is clear from this equation that one may define the refined current
\begin{equation}
    J'_e=\frac{1}{g_5^2} \hodge \dd C - \frac{\kappa}{g_5 e_A}\hodge\dd A \, ,
\label{eq:refinedmixcurrent}
\end{equation}
which is conserved. This means that a kinetic mixing term~\eqref{eq:kineticmixing} cannot generate a potential for the axion. This expectation even holds true when the one-form symmetry associated with $A$ is badly broken since the equation of motion for $C$ always sets the exterior derivative of eq.~\eqref{eq:refinedmixcurrent} to zero~\footnote{We thank Matt Reece for emphasizing this point.}.

The action~\eqref{eq:kineticmixing} in fact has a joint conserved $U(1)\times U(1)$ one-form symmetry, even when $\kappa \neq 0$. Coupling $C$ and/or $A$ to gluons partially breaks this symmetry to a $U(1)\times \mathbb{Z}_N$ subgroup, corresponding to simultaneous appropriately scaled shifts in both gauge fields (and a remnant discrete shift symmetry that depends on integer coefficients in front of the gluon coupling terms). Leveraging this fact, one can always do a field redefinition to go to a basis in which \textit{only} $C$ is coupled to gluons. In that sense, these kinetically mixed models only have one true QCD axion. A caveat is that if we include additional Chern-Simons terms in our action, such as couplings between $C$, $A$, and Standard Model photons, there is generally not a basis in which all couplings are delegated entirely to $C$, so both $A$ and $C$ will have Chern-Simons terms. From the one-form symmetry point of view, this is a manifestation of the fact that the subgroup the leaves the gluon couplings invariant and the subgroups that leave the other Chern-Simons terms invariant need not be the same. Finally, we note that one can also get rid of the kinetic mixing altogether via a field redefinition, although this obscures flux quantization. Models with these types of bases are nevertheless of great phenomenological interest since they can be used to obtain non-standard effective 
axion coupling terms~\cite{Babu_1994,Agrawal:2017cmd,Dror:2020zru}.

We may also add irrelevant operators to our 5D action. For example, consider an Euler-Heisenberg type operator of the form
\begin{equation}
    S^\mathrm{5D}=\int \left(-\frac{1}{2g_5^2}\dd C \wedge \hodge \dd C - \mathcal{O}\left[\frac{(\dd C)^4}{\Lambda^4}\right]\right) \, ,
\end{equation}
where $\Lambda$ is some arbitrary energy scale ensuring that the operator has correct mass unit. An operator of this type arises by e.g. integrating out the heavy particle states that we showed yield an axion potential in Section~\ref{sec:electriccharge}. The presence of these irrelevant operators modifies the electric one-form symmetry via the equation of motion
\begin{equation}
\dd \left(J_e + \mathcal{O}\left[\frac{(\dd C)^3}{\Lambda^4}\right]\right)=0 \, .
\label{eq:irrelevant}
\end{equation}
However, it is obvious from eq.~\eqref{eq:irrelevant} that we may define a refined one-form symmetry current
\begin{equation}
    J_e' = \frac{1}{g_5^2}\hodge \dd C+\mathcal{O}\left[\frac{(\dd C)^3}{\Lambda^4}\right] \, ,
\end{equation}
which is conserved. Operators of this form hence do not pose a danger to the quality of the axion. This should be contrasted with the breaking of zero-form symmetries by irrelevant operators where a refined current can rarely be defined. This once again illustrates the fact that long range local physics remains largely unaffected by one-form symmetry breaking. Effects from the heavy states are exponentially suppressed and only show up by dimensionally reducing the full UV theory~\cite{McGreevy:2022oyu}.

Naturally, one can add other irrelevant operators consisting of non-derivatively coupled fields to the action~\eqref{eq:axiongluonaction}, such as irrelevant operators involving multiple $C$-covariant derivatives and new fields. These operators generally break the electric $U(1)_e^{(1)}$ one-form symmetry with an appropriate scale suppression on the right hand side of the electric one-form current equation, $\dd J_e = \frac{1}{\Lambda^n}\mathcal{O}$ where $n\geq 1$. These operators will hence generally have a lesser effect on the axion potential than the marginal operators discussed thus far. We also emphasize a key distinction between the PQ and the extra-dimensional axion scenario in this regard: One-form symmetry breaking, which is mandated by basic quantum gravity expectations, can be entirely achieved by the heavy charged states discussed in Section~\ref{sec:electriccharge}. We saw that these states have an exponentially suppressed effect on the axion potential and otherwise have no one-form symmetry-breaking trace once integrated out. With these heavy states added to~\eqref{eq:axiongluonaction}, the few other symmetry-breaking irrelevant operators one can write down are not a generic quantum gravity expectation like they are for the PQ symmetry of the PQ axion (see Section~\ref{sec:gravitationalconsiderations} for further discussion on gravitational expectations).

\section{Magnetic Charges}\label{sec:magneticbreaking}
The free one-form gauge field $C$ possesses a magnetic $U(1)_m^{(2)}$ two-form symmetry in addition to its electric $U(1)_e^{(1)}$ one-form symmetry. While the magnetic symmetry plays no direct role in protecting the axion potential, magnetic symmetry breaking can accommodate new physics that leads to breaking of the electric one-form symmetry. In this section, we discuss how monopoles charged under any $U(1)$ gauge field generate an axion potential provided that the 5D action includes a Chern-Simons term. We also briefly discuss how monopoles charged directly under $C$ -- and other branes with worldvolume degrees of freedom -- affect the electric one-form symmetry.

We first study what happens to the electric one-form symmetry of $C$ when we add a coupling to monopoles of another $U(1)$ one-form gauge field $A$. Suppose we add the following term to our 5D axion action
\begin{equation}
    S^\mathrm{5D}\supset \int\left( -\frac{1}{2 e_A^2}\dd A \wedge\hodge \dd A+\frac{1}{8\pi^2} C\wedge \dd A \wedge \dd A\right),
\label{eq:monopoleanomalyterm}
\end{equation}
where $e_A$ is the gauge coupling of $A$ with mass dimension $[e_A]=-1/2$. Here, we have chosen the integer prefactor $1$ in the coupling term for simplicity. The action~\eqref{eq:monopoleanomalyterm} is the $U(1)$ analogue of the $SU(N)$ action considered in Section~\ref{sec:abjanomaly}. Like in the $SU(N)$ case, the electric $U(1)_e^{(1)}$ one-form symmetry is broken completely. The current equation is
\begin{equation}
    \dd J_e = \frac{1}{8\pi^2}\dd A \wedge \dd A \, .
\label{eq:monopoleabjbreaking}
\end{equation}
While the symmetry is broken, it is worth noting that a non-invertible symmetry remains, as studied in ref.~\cite{Damia:2022bcd}. We saw that a term of this form with a non-Abelian gauge field leads to an axion potential via gauge instantons in the infrared. There are no $U(1)$ instantons on a topologically trivial 4D spacetime~\cite{Srednicki:2007qs,Shifman:2012zz}, so one would naively be led to believe that the term~\eqref{eq:monopoleanomalyterm} cannot generate a potential for the axion. However, it was found by ref.~\cite{Fan:2021ntg} that in 4D axion theories with a coupling of the form $\theta \dd A \wedge \dd A$, virtual monopoles magnetically charged under $A$ do induce an effective potential for the axion~\footnote{Ref.~\cite{Cordova:2022ieu} linked this effective potential in 4D to the further breakdown of the axion shift symmetry, which would have a surviving non-invertible symmetry remnant were monopoles not included.}. This can also happen in 5D theories with a term of the form~\eqref{eq:monopoleanomalyterm}, which we now outline.

In 5D, monopoles are strings rather than point particles. Monopole strings naturally arise in theories where the $U(1)$ gauge field descends from a Higgsed $SU(N)$ gauge field as first found by 't Hooft and Polyakov~\cite{tHooft:1974kcl,Polyakov:1974ek} in 4D. The 5D counterpart of 't Hooft-Polyakov solutions have been studied analytically in the BPS limit~\cite{Tong:2005un,Blanco-Pillado:2006gsr,Boyarsky_2002}. The static 5D monopole solution allows for two types of perturbations. The first is transverse waves traveling up and down the string. The second is a type of longitudinal wave, which crucially carries electric charge. This latter type of perturbation is the 5D analogue of the dyonic excitations of the standard 't Hooft-Polyakov monopole~\cite{Julia:1975ff,Shifman:2012zz}.

In the BPS limit, these electric excitations are described by the 5D monopole worldsheet action
\begin{equation}
    S^{\mathrm{Worldsheet}}=\int\left( -\frac{1}{2}\frac{T_M}{m_W^2}\dd_A \sigma \wedge \hodge \dd_A \sigma + \frac{1}{2\pi} C \wedge \dd_A \sigma \right) \, ,
\label{eq:monopoleworldsheetaction}
\end{equation}
where $\sigma$ is a $2\pi$ periodic scalar, $\dd_A \equiv \dd + 2A$ is an $A$-covariant derivative, and where $m_W$ and $T_M$ are the $W$ boson mass and monopole string tension, respectively. One can understand why these worldsheet degrees of freedom must arise via higher-form symmetry gauging~\cite{Heidenreich:2020pkc} and anomaly inflow~\cite{Choi:2022fgx,Fukuda:2020imw}. They were also recently explicitly derived in 4D in ref.~\cite{Garcia-Valdecasas:2024cqn}.

Wrapping the monopole worldsheet around the $S^1$ and Fourier expanding the field content of~\eqref{eq:monopoleworldsheetaction}, the worldsheet theory reduces to a 4D monopole worldline action of the form studied in ref.~\cite{Fan:2021ntg}, plus couplings between higher KK modes on the worldsheet not involving the axion $\theta$. One would then be tempted to conclude that the axion gets an induced potential of the form~\cite{Fan:2021ntg}
\begin{multline}
    V(\theta)=-\sum_{\ell=1}^\infty \frac{R^2 T_M^2 m_W^2}{8\pi^2 \ell^3}e^{-4\pi^2 \ell R T_M/m_W}\cos{(\ell \theta)}\\
    \times\left(1+\frac{3m_W}{4\pi^2 \ell R T_M}+\frac{3 m_W^2}{16\pi^4\ell^2 R^2 T_M^2}\right) \, ,
\label{eq:monopolepotential}
\end{multline}
from closed monopole worldsheets, akin to the potential one would obtain if the monopole solution only existed in 4D (which would occur if $m_W\lesssim 1/R$). However, we caution that this potential should be taken as approximate. It does not account for Coulomb interactions, and is subject to screening whenever light charged fermions are present in the theory~\cite{Fan:2021ntg}. Furthermore, in viewing the wrapped worldsheet as an instanton with a two-torus topology, there may be additional terms from ``twisting'' the worldsheet along the $S^1$ direction that are not accounted for in eq.~\eqref{eq:monopolepotential}. 

It is clear from the instanton-like electric one-form symmetry breaking~\eqref{eq:monopoleabjbreaking} that the monopole worldsheet induced axion potential arises in a similar fashion to the potential generated by $SU(N)$ gauge instantons. It is also noteworthy that the potential~\eqref{eq:monopolepotential} is of the same form as the potential found in Section~\ref{sec:abjanomaly}. The connection between closed monopole worldvolumes and gauge instantons in Higgsed $SU(N)$ theories -- and some of the related open questions posed above -- deserve further illumination and are the subject of ongoing study~\cite{GarciaGarcia:future}.

Finally, we remark that if the monopole is charged directly under $C$, a Chern-Simons term of the form $C\wedge \dd C \wedge \dd C$ still gives rise to an axion potential similar to~\eqref{eq:monopolepotential}. However, for a $C$ monopole, even if we set the Chern-Simons term to zero, electric modes on the monopole string persist. This means that the electric one-form symmetry is still broken, and the current equation is locally modified to $\dd J_e = j_{\mathrm{electric\; modes}}$ on the monopole worldsheet. This suggests that monopoles -- and branes in general -- with localized degrees of freedom that break the $U(1)_e^{(1)}$ symmetry can still have an effect on the axion potential. We leave further study of these branes for the future.

\section{Gravitational Considerations}\label{sec:gravitationalconsiderations}

The axion quality problem is often framed in terms of global symmetry-breaking effects from quantum gravity. Although the effects we have considered in our 5D toy model are all field-theoretic in nature, they may be used to infer the size of expected corrections to the extra-dimensional axion potential from quantum gravity via various forms of the Weak Gravity Conjecture (WGC)~\cite{Arkani-Hamed:2006emk,delaFuente:2014aca, Heidenreich:2015wga, Cordova:2022rer}. While many of the WGC-induced connections between the extra-dimensional axion quality problem and quantum gravity expectations have been drawn elsewhere, they dovetail nicely with the present classification. Similarly, the Completeness Hypothesis imposes requirements on the spectrum of states in our theory, and these requirements in turn determine how the electric and magnetic higher-form symmetries associated with the axion gauge field are broken.

\subsection{5D Electric WGC}
The 5D electric Weak Gravity Conjecture requires a particle in the theory with mass $m_{\rm 5D}$ and charge $q$ satisfying
\begin{align}
    m_{\rm 5D} \lesssim  g_5 q M_{\rm Pl, 5D }^{3/2} \, ,
\end{align}
where $M_{\rm Pl, 5D }$ is the 5D Planck scale. This sets an expectation for the minimum size of corrections to the axion potential arising from electrically charged matter, since the contributions to eq.~\eqref{eq:chargedpot} from the WGC particle scale exponentially with
\begin{align} \label{eq:Sewgc}
S = 2 \pi m_{\rm 5D} R \lesssim 2 \pi q g_5 R M_{\rm Pl,5D}^{3/2} = q M_{\rm Pl} / f \, ,
 \end{align}
 where we have used $M_{\rm Pl}^2 = 2 \pi R M_{\rm Pl, 5D}^3$. This exemplifies the statement that the quality problem for extra-dimensional axions is in general the logarithm of the usual 4D quality problem, which is polynomial in $f/M_{\rm Pl}$. Under dimensional reduction, the 5D electric WGC reduces to the usual 4D electric WGC for the massless vector (if the vector zero mode survives), while eq.~\eqref{eq:Sewgc} recapitulates the general 4D axion WGC, $f \lesssim \frac{q}{S} M_{\rm Pl}$.

\subsection{5D Magnetic WGC}

The 5D magnetic WGC posits that the local electric description of a 5D $U(1)$ should break down at a scale
\begin{align}
\Lambda \lesssim g_{5} (M_{\rm Pl,5D})^{3/2} \, ,
\end{align}
associated with magnetically charged degrees of freedom. There are various ways to parse this bound in the current setting. Additionally requiring $\Lambda > 1/R$ (i.e., that there be a sensible regime in which the 5D description holds) leads to the parametric form of the 4D axion magnetic WGC, $f \lesssim M_{\rm Pl}$~\cite{delaFuente:2014aca}. With the same restriction, the 5D magnetic WGC also implies that the minimal corrections to the axion potential from magnetic monopole strings in the presence of a $U(1)$ ABJ anomaly, as given by eq.~\eqref{eq:monopolepotential}, scale exponentially with
\begin{align}
S \propto \frac{R}{g_5^2} \lesssim M_{\rm Pl}^2 R^2 \, .
\end{align}
The expected contribution of monopole strings to the extra-dimensional axion quality problem is thus doubly exponentially suppressed.

\subsection{2-form WGC}

Gauging the electric one-form symmetry as in Section~\ref{sec:gaugingelec} forbids terms that explicitly break the symmetry protecting the extra-dimensional axion, at the cost of the axion being eaten to form a longitudinal mode whose mass is proportional to $e_B$. This naively raises the possibility of protecting a light axion with an unreasonably good global symmetry by taking $e_B$ arbitrarily small. However, the 2-form WGC~\cite{Heidenreich:2015nta} obstructs this by requiring the existence of 1D charged objects whose charge $q_2$ and tension $T_2$ satisfy
\begin{align}
    \frac{2}{3} T_2^2 \leq q_2^2 e_B^2 M_{\rm Pl, 5D}^3 \, .
\end{align}
Phrased in terms of the mass $m_{\tilde A}$ of the 4D gauge field, the 2-form WGC implies
\begin{align}
m_{\tilde A} \gtrsim \frac{k}{q_2} \frac{T_2}{g_4 M_{\rm Pl}} \, ,
\end{align}
so that the vector mass may not be made arbitrarily small without introducing dynamical charged strings into the low-energy theory. 

\subsection{Completeness Hypothesis}
In its most basic form, the Completeness Hypothesis postulates that all charges consistent with Dirac quantization must be realized by physical states in a gauge theory that is consistently coupled to gravity~\cite{Polchinski_2004,Palti:2019pca}. For the extra-dimensional axion of the form discussed in this work, the Completeness Hypothesis demands the existence of unit electric and unit magnetic charge states (but unlike the weak gravity conjecture, no restriction is placed on the masses of these states). Thus, both the $U(1)_e^{(1)}$ and $U(1)_m^{(2)}$ symmetries must be broken completely, without the possibility of a surviving remnant discrete symmetry. This generally leads to $2\pi$ periodic potentials of the form discussed in Sec.~\ref{sec:electriccharge} and~\ref{sec:magneticbreaking}~\footnote{We note that the Completeness Hypothesis has been studied and generalized in the context of higher-form gauging and non-invertible symmetries, where the notion of completeness of the spectrum can be shown to be equivalent to the endability of extended operators~\cite{Heidenreich:2021xpr}.}.

\section{Conclusion and Outlook} \label{sec:conclusion}
In this work, we have made explicit the sense in which an axion descending from an extra-dimensional one-form gauge field is protected by an electric one-form symmetry. This protection mitigates harmful axion potential corrections from UV physics. This higher-form symmetries perspective has a number of advantages over the conventional statement that it is gauge invariance that protects the extra-dimensional axion. First, it is more physical since gauge symmetry is a redundancy of description and not a physical symmetry. Second, it is on closer footing with the standard understanding of how PQ symmetry protects the potential of the PQ axion. Third, it is much more powerful in identifying and understanding new physics that induces a quality problem.

In accordance with this third point, we have identified three ways of breaking the protective electric one-form symmetry down to a discrete one-form symmetry: adding electrically charged matter to the theory, gauging a magnetic two-form symmetry, and turning on an ABJ anomaly. We have shown that all three types of breaking result in an axion potential with features dictated by surviving discrete one-form symmetries. We have also studied what happens when the electric one-form symmetry is gauged. In all four cases, we have discussed what each symmetry modification means for the axion quality problem.

We have also studied how the presence of magnetic monopoles with dyon collective coordinates affects the electric one-form symmetry and in turn the axion potential. If the 5D action has a $U(1)$ ABJ anomaly term and monopoles, this gives rise to a potential via closed Euclidean monopole worldvolumes. We discuss how these monopole contributions fall under the three types of breaking discussed previously. However, these closed worldvolumes and their relation to gauge instantons should be further illuminated and are the subject of ongoing work~\cite{GarciaGarcia:future}. Furthermore, it would be interesting to understand more generally how branes with local degrees of freedom that break the electric one-form symmetry affect the axion from the higher-form symmetries point of view. We leave this question -- and many rich details on the connection between extra-dimensional monopoles, generalized symmetry breaking, and topology -- for future study.

Additionally, we have discussed how generalized weak gravity conjectures and the Completeness Hypothesis constrain the parameters involved in electric one-form symmetry breaking in the 5D actions considered. It would be interesting to study more detailed quantum gravity inspired models with a more complex topology from the higher-form symmetry breaking point of view. Moreover, the one-form electric symmetry breaking discussed in this work is interpretable in a dual frame where it is the magnetic higher-form symmetry of the dual axion Kalb-Ramond field that is broken. We leave the study of these more complicated models and the magnetic dual for future endeavors.

Finally, we note that this work presented a powerful use case for generalized symmetries in understanding and addressing the quality problem, which is of central importance to particle phenomenology. We believe that similar logic can applied to other phenomenological models involving dimensional reduction, or more generally, to any model where an important symmetry in the infrared descends from a generalized symmetry in the ultraviolet. We hope that this work inspires efforts to attempt to find more significant use cases for generalized symmetries in particle phenomenology more broadly.

\acknowledgements We thank Clifford P. Burgess, Ananth Malladi, and Dan Sehayek for useful discussions, Prateek Agrawal, Isabel Garcia Garcia, Matthew B. Reece, and Ken Van Tilburg for useful comments on the manuscript, and Stephen Shenker for asking good questions. NC is supported in part by the U.S. Department of Energy under the grant DE-SC0011702 and gratefully acknowledges the hospitality of the Kavli Institute for Theoretical Physics (KITP), supported by the National Science Foundation under Grant No.~NSF PHY-1748958. MK is supported by a James Arthur Graduate Assistantship at New York University. MK was a KITP graduate fellow when most of this work was carried out. The fellowship was supported in part by the Heising-Simons Foundation, the Simons Foundation, and grant no.~PHY-2309135 to KITP. This material is based upon work supported by Grant No.~NSF PHY-2210551.

\bibliography{5d_quality}

\end{document}